\documentclass[journal=jacsat,manuscript=article]{achemso}
\usepackage[version=3]{mhchem} 
\usepackage{siunitx} 
\usepackage{soul}

\author{S.G. Bishop}
\affiliation{School of Engineering, Cardiff University, Queen's Building, The Parade, Cardiff, UK, CF24 3AA}
\author{J.K. Cannon}
\affiliation{School of Engineering, Cardiff University, Queen's Building, The Parade, Cardiff, UK, CF24 3AA}
\author{H.B. Ya\u{g}c{\i}}
\affiliation{School of Engineering, Cardiff University, Queen's Building, The Parade, Cardiff, UK, CF24 3AA}
\author{R.N. Clark}
\affiliation{School of Engineering, Cardiff University, Queen's Building, The Parade, Cardiff, UK, CF24 3AA}
\author{J.P. Hadden}
\affiliation{School of Engineering, Cardiff University, Queen's Building, The Parade, Cardiff, UK, CF24 3AA}
\author{W.W. Langbein}
\affiliation{School of Physics and Astronomy, Cardiff University, Queen's Building, The Parade, Cardiff, UK, CF24 3AA}
\author{A.J. Bennett}
\affiliation{School of Engineering, Cardiff University, Queen's Building, The Parade, Cardiff, UK, CF24 3AA}
\alsoaffiliation{School of Physics and Astronomy, Cardiff University, Queen's Building, The Parade, Cardiff, UK, CF24 3AA}
\email{BennettA19@cardiff.ac.uk}


\title[Evanescent-Field Assisted Photon Collection from Quantum Emitters under a Solid Immersion Lens]
  {Evanescent-Field Assisted Photon Collection from Quantum Emitters under a Solid Immersion Lens}

\newcommand{\ignore}[1]{}

\begin{document}

\begin{abstract}
Solid-state quantum light sources are being intensively investigated for applications in quantum technology. A key challenge is to extract light from host materials with high refractive index, where efficiency is limited by refraction and total internal reflection. Here we show that an index-matched solid immersion lens can, if placed sufficiently close to the semiconductor, extract light coupled through the evanescent field at the surface. \ignore{This enables collection of light over large numerical apertures. In an experiment we automate the selection of a representative sample of colour centres from an ensemble in aluminum nitride, and study how changing the spacer between the semiconductor and the lens changes the collection of this evanescent field.}Using both numerical simulations and experiments, we investigate how changing the thickness of the spacer between the semiconductor and first lens impacts the collection efficiency. Using automatic selection and measurement of 100s of individually addressable colour centres in several aluminium nitride samples we demonstrate spacer-thickness dependent photon collection efficiency enhancement, with a mean enhancement of a factor of $\SI{4.2}{}$ and a highest measured photon detection rate of $\SI{743\pm4}{kcps}$. \ignore{In the limit of vanishing thickness, our scheme can collect \SI{37}{\percent} of all emitted photons from an in-plane dipole point source into 0.9 numerical aperture.} 
\end{abstract}


\section{Introduction}
The number of photons collected from a source is of paramount importance for any quantum optoelectronic technology. This is especially critical for quantum light sources such as colour centres, quantum dots and single ions where increasing the photon flux is necessary to allow measurement of correlation effects. However, in solid state systems these emitters are embedded in high refractive index materials such as diamond (index 2.4 @ \SI{600}{\nano\meter}), gallium arsenide (3.4), silicon carbide (2.6) or indium phosphide (3.5). Refraction at the air-dielectric surface reduces the solid angle over which light may be collected inside the material. \ignore{At high angles emission is totally internally reflected at the surface, placing an upper limit on the efficiency.} At high emission angles, light is lost at the interface due to total internal reflection (TIR), placing an upper limit on the efficiency. Various micro-processed architectures have been demonstrated to increase collection efficiency (CE), including etched micro-pillars\cite{Santori2002, Somaschi2016}, nanowire antennae \cite{Babinec2010, Claudon2010}, photonic crystals\cite{Noda2007,Lodahl2015a}, dielectric resonators \cite{Kuznetsov2016,Mignuzzi2019}, and free-space cavities\cite{Tomm2021, Dolan2010}. All require the photonic structure to be aligned to the emitter in space and frequency, which does not easily support the study of randomly positioned, broadband emitters. In contrast, a millimeter-sized solid immersion lens (SIL) \cite{Born1999, A.N.Vamivakas2007, Chen2018},\ignore{consisting of hemispherical dielectric} a dielectric half-sphere, offers an inherently wide-field CE enhancement over a broad spectral range. This enhancement is optimised for buried emitters when the refractive index of the SIL is matched to the index of the semiconductor \cite{Barnes2002,Serrels2008}.

Quantum dots in gallium nitride are unique in offering quantum light emission in the blue and UV part of the spectrum \cite{Zhu2016a}. Due to the recent discovery of quantum emitters in gallium nitride (GaN)\cite{Berhane2017,Zhou2018} and aluminium nitride (AlN)\cite{Bishop2020}, there is renewed interest in nitrides as a platform for quantum optics in the visible and near infra-red. Polarised quantum light emission at room temperature, coupled with established processing technology, presents an exciting platform combining optoelectronic components, integrated photonics, and quantum emitters\cite{Wan2020}. In addition, the III-Nitrides present one of the only known room temperature quantum emitters in the telecom band \cite{Zhou2018}. 

In this report, we demonstrate how a SIL enhances the CE from colour centres in AlN \cite{Bishop2020}. We use confocal microscopy with an index matched ZrO\textsubscript{2} hemispherical SIL (h-SIL) ($n_{\mathrm{SIL}}=2.16$ at  \SI{600}{\nano\meter}) on AlN templates ($n_{\mathrm{AlN}}=2.15$ at  \SI{600}{\nano\meter}). We engineer the thickness of a layer of PMMA ($n_{\mathrm{PMMA}}=1.49$ at \SI{600}{\nano\meter}) to avoid the formation of an air gap between the semiconductor and SIL. Using finite difference time domain (FDTD) simulations, we show that engineering the thickness of the polymer is important for increasing the CE with a high NA lens due to evanescent coupling through the PMMA layer. 

\section{Methods}
\subsection{An ideal hemispherical SIL}

\begin{figure}[t]
    \centering
    \includegraphics{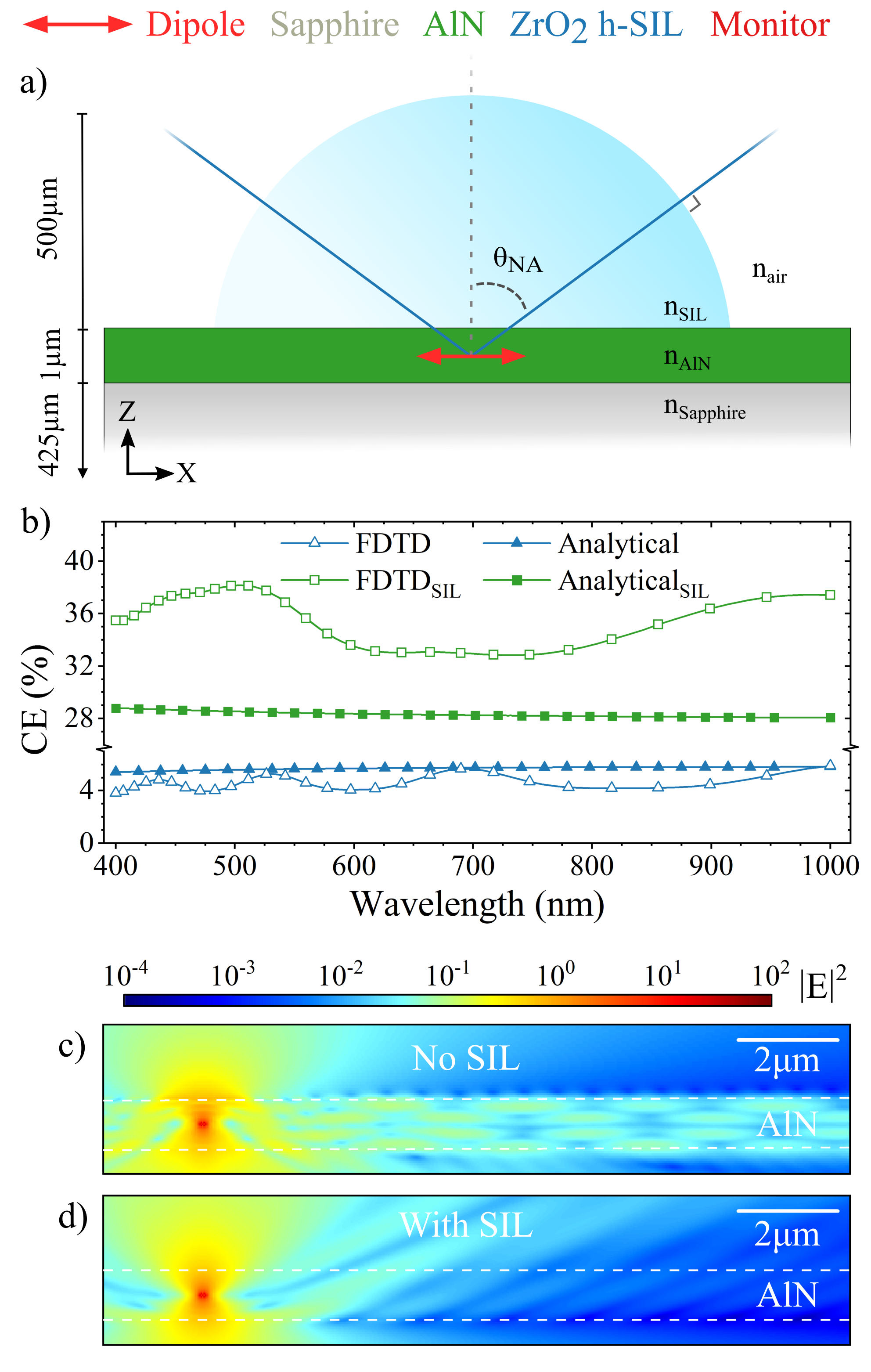}
    \caption{Hemispherical solid immersion lens. a) Illustration of the geometry. b) Calculated (filled symbols) and simulated (empty symbols) collection efficiency (CE) as a function of the wavelength of the dipole with (green) and without (blue) a hemispherical SIL. c,d) XZ slices of the log of the electric field across the dipole without and with the SIL respectively.}
    \label{fig:intro}
\end{figure}

The hemispherical SIL, a truncated sphere bisected across its origin, exploits the SIL's geometry to increase photon CE by limiting refraction at the semiconductor-SIL interface by reducing the index difference. Additionally, aplanatic imaging is achieved due to the normal incidence of light at the SIL-air interface as shown in Fig.\ref{fig:intro}(a). The NA of the imaging system is therefore increased by the refractive index within the semiconductor, reducing the axial and lateral size of both the excitation and collection point-spread functions by a factor of the SIL refractive index and the square of the index, respectively, relative to imaging through a flat interface \cite{Bishop2022}.

The CE may be analytically estimated from Barnes $et$ $al$ \cite{Barnes2002} considering a dipole orientated in the plane of an infinite AlN slab\cite{Bishop2020}. The blue and green plots with filled symbols in Fig. \ref{fig:intro}(b) demonstrate the analytically determined CE without and with a ZrO\textsubscript{2} SIL, respectively. The reflection at the SIL-air interface is accounted for with Fresnel equations, with a reflectivity of 13\%. To take into account the effect of other reflections, such as that from the AlN to sapphire interface, which are not included in the analytical analysis, we perform Finite Difference Time Domain (FDTD) simulations in Lumerical. We simulate the AlN layer surrounded by a sapphire substrate ($n=1.7$) and a homogeneous ZrO\textsubscript{2} SIL using outgoing boundary conditions (perfectly matched layers). Anti-symmetric and symmetric boundary conditions are used across the dipole to reduce the simulated region. \ignore{To avoid simulating the macroscopic SIL, we include a ZrO\textsubscript{2} homogeneous dielectric layer above the AlN. The electric field profile is then calculated at planes within the ZrO\textsubscript{2} layer,} The loss due to the reflection at the SIL-air interface is included separately. The FDTD simulation predicts a wavelength-varying and enhanced CE with regards to the analytically determined value, due to the finite reflection from the AlN-sapphire interface, shown as the blue and green lines in Fig.\ref{fig:intro}(b) with empty symbols. This is evident from the electric field cross-section plots through the dipole in Fig. \ref{fig:intro}(c-d). In this scenario, the expected enhancement of light collection into a NA of 0.9 is increased by a factor of approximately 8 by the SIL, to a mean value of \SI{35}{\percent} across the visible spectrum.

\subsection{Ensemble scanning}
\begin{figure}[ht]
    \centering
    \includegraphics{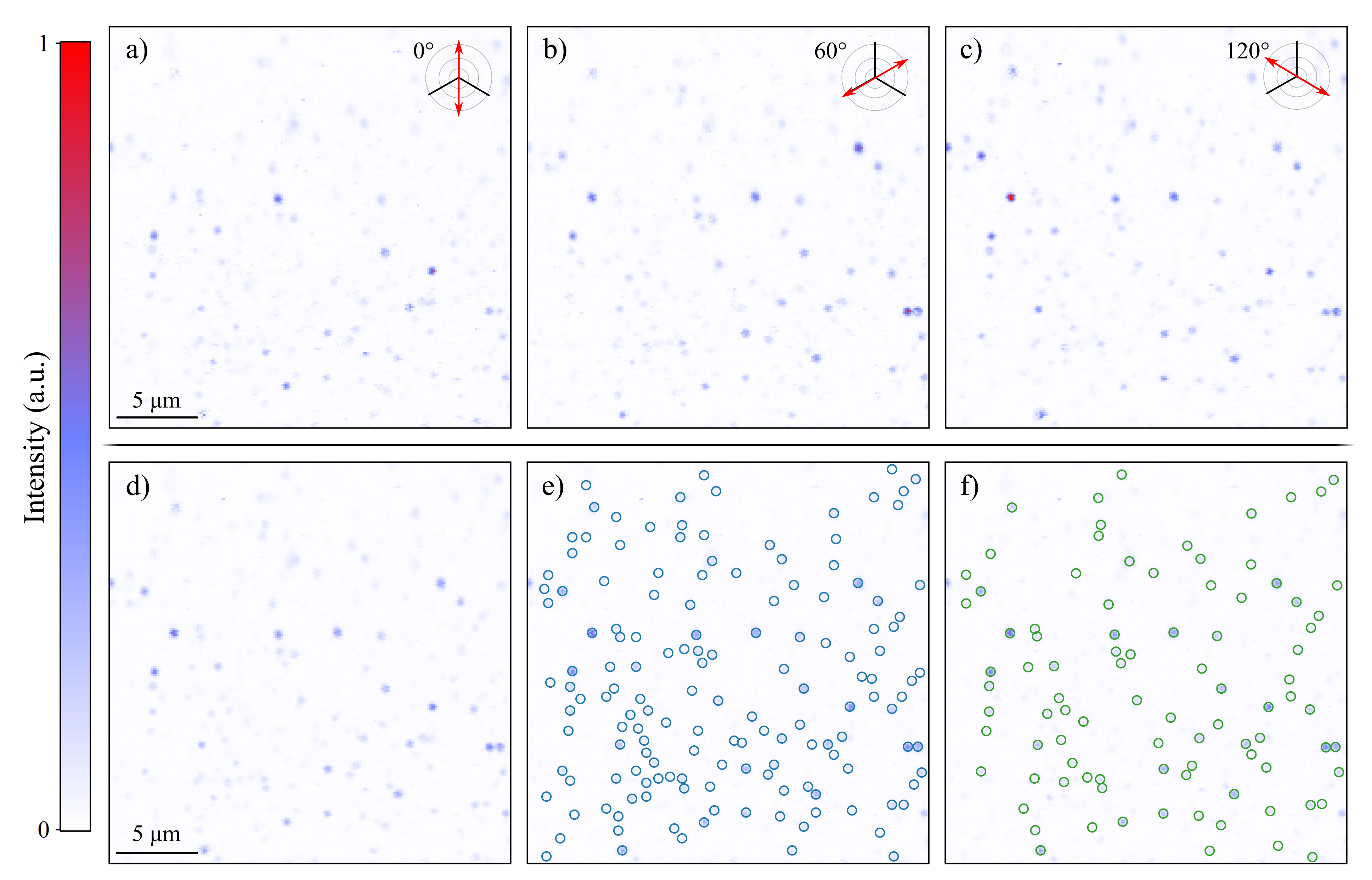}
    \caption{Automated selection of emitters in confocal scan maps. a,b,c) Excitation polarisation resolved confocal scans, where the excitation laser polarisation is rotated in the plane of the sample at an angle of \SI{0}{\degree}(a), \SI{60}{\degree}(b) and \SI{120}{\degree}(c) respectively. d) Combined scan map where each pixel represents the mean of the 3 polarisation resolved scan maps. e) Emitter locations determined with the fast peak finding technique. d) Corresponding filtered emitter locations after Gaussian fitting to each emitter's point-spread function.}
    \label{fig:preselection}
\end{figure}

To experimentally quantify the enhancement in CE, we use statistical analysis over an ensemble of individually addressable emitters. To reduce bias in the selection of the emitters, we developed an automated routine to locate and characterise emitters. Initially, confocal sample maps were measured revealing the presence of emitters in the AlN epilayer using an optical dual-axis 4f scanning system \cite{Bishop2020}. Optical filtering was used to measure only the \SI{550}{} to \SI{650}{\nano\meter} spectral range, suppressing fluorescence from chromium impurities in the sapphire substrate. Due to the linearly polarised nature of the excitation laser at \SI{532}{\nano\meter} and the fact that each emitter has a linearly-polarised excitation dipole with unknown orientation \cite{Bishop2020}, three scans maps were taken with the laser's linear polarisation angle at \SI{0}{\degree}, \SI{60}{\degree} and \SI{120}{\degree}, shown in Fig.\ref{fig:preselection}(a-c). We combine these scan maps in (d), which is the mean of scans (a-c), to avoid pre-selecting emitters aligned to the laser polarisation. \ignore{We conclude from this data (analysis not included here) that there is no preferred orientation(s) of the the emitters with respect to the crystallographic axes \cite{Cannon2022}.} As reported in our previous work, AlN colour centres in this samples have a zero-phonon line at \SI{595}{\nano\meter} with a broad phonon side-band extending to \SI{700}{\nano\meter} \cite{Bishop2020}. 
 
The resultant scan map was then fed into an algorithm that locates emitters in the following manner. Initially, the algorithm removes single pixel noise by using a $3\times3$ pixel median filter and further smooths the image by convolving the image with a two-dimensional Gaussian function. A fast pixel-by-pixel analysis then locates potential emitter locations based on the intensity of neighbouring pixels. The 126 candidate emitter locations are marked in Fig.\ref{fig:preselection}(e). It can be observed in the image that the algorithm overestimates the number of emitters and occasionally selects features smaller than the diffraction limited lateral spot size. Therefore, a 2-dimensional Gaussian function is fit to each emitter. The Gaussian fit parameters are also used to rank the emitters based on the Gaussian width being close to the expected diffraction limited point-spread function of the microscope, symmetry in X and Y, amplitude and goodness of fit $R^2$ value.  This Gaussian fit identifies the centre of the emitters' point spread function with greater precision than the scan step-size. Post filtering, 92 emitters are identified in the scan map Fig.\ref{fig:preselection}(f).

Power dependent saturation measurements are then taken on each candidate emitter sequentially using the following automated process. Firstly, the scanning system moves to the approximate emitter location. X,Y and Z line scans are performed to find the maximum along each axis. The linear polarisation of the \SI{532}{\nano\meter} excitation laser is rotated to maximally couple to the emitter's absorption dipole. Then a laser-power $P$ dependent intensity $I(P)$ was measured and used to quantify the fitted saturation intensity at infinite power $I_{\infty}$ as defined by the fit function $I(P) = I_{\infty}P/(P+P_{sat})$. Measurements with a saturation power greater than $\SI{2}{\milli\watt}$ are removed from each data-set as this is a good proxy for linear power dependent behaviour, which is not consistent with a single colour centre (this occurs with less than \SI{5}{\percent} frequency). No polarisation optics were included in the collection path of our microscope. Throughout the measurements, photo-bleaching was not observed for any emitters but random telegraph noise in the detected count rate is common. 

\section{Results and Discussion}

\subsection{Collection efficiency enhancement}
Using the automated system we set out to determine the efficiency enhancement for an ensemble of emitters under a SIL. We emphasise that the presence of an air-gap between the SIL and AlN suppresses such an enhancement. An air gap can arise from a non-flat back surface of the SIL and/or a non-flat surface of the semiconductor. To prevent such a gap, we add a thin layer of a polymer, PMMA, which is soft enough to fill the gap between the SIL and AlN adapting to their surfaces, in line with other reports \cite{A.N.Vamivakas2007, Bishop2022}. Three samples of AlN-on-sapphire from the same wafer were studied. One with no SIL (SP1), one with \SI{212}{\nano\meter} of PMMA between the SIL and AlN (SP2) and one with \SI{70}{\nano\meter} of PMMA (SP3). The PMMA thickness was determined using an ellipsometer prior to the addition of the SIL. 

\begin{figure}[ht]
    \centering
    \includegraphics{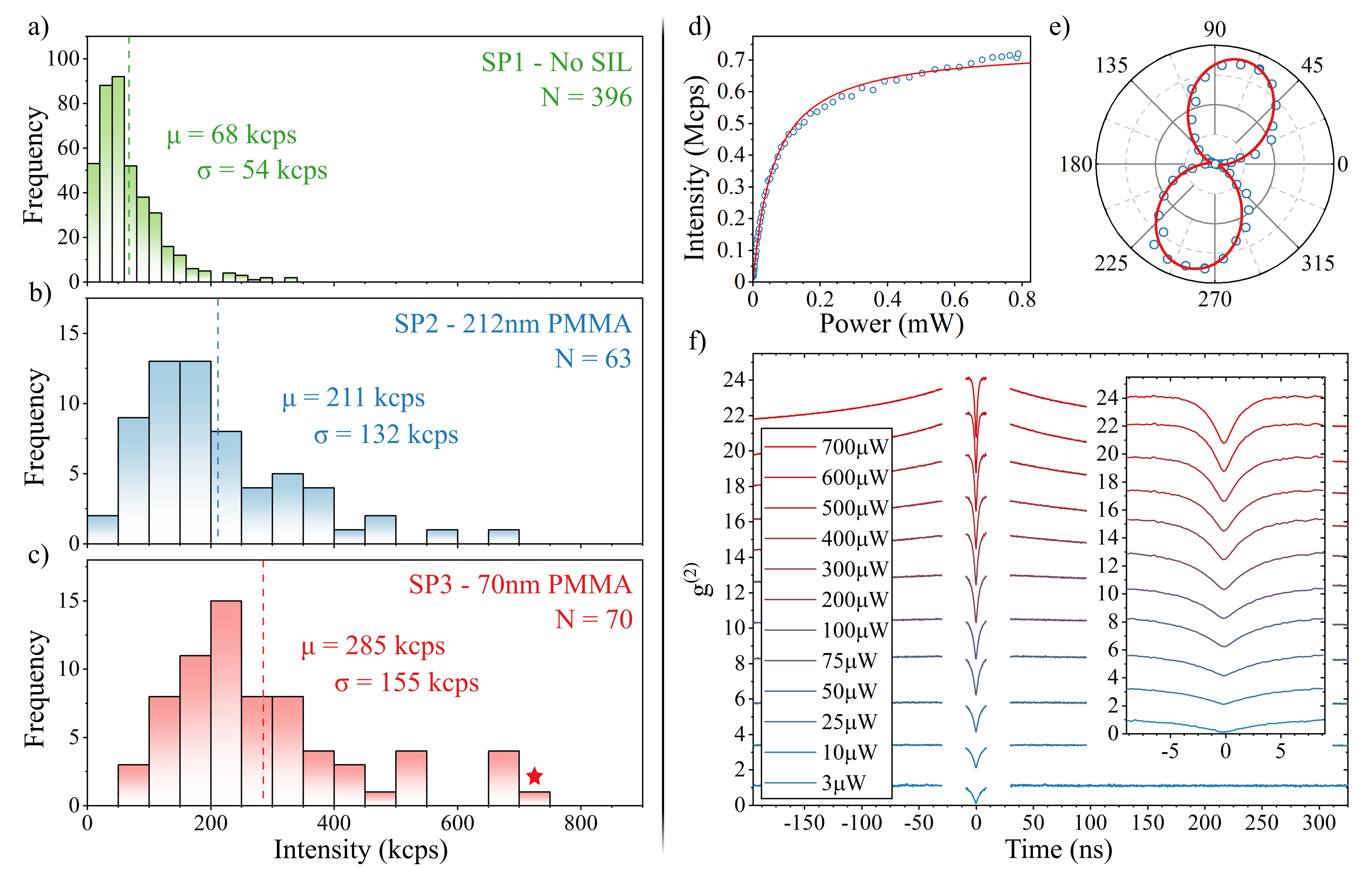}
    \caption{Evanescent-field assisted collection by a hemispherical SIL. a-c) Histogram representing the fitted saturation intensity at infinite power power, $I_{\infty}$, for a number of emitters in samples SP1, SP2 and SP3. The polymer thickness is labelled on each plot. For the brightest emitter in SP3 (marked with a red star) we show the power-dependent intensity saturation curve (d), the intensity as a function of excitation polarisation (e) and power dependent auto-correlation histograms (f). }
    \label{fig:stats}
\end{figure}

The statistical distribution of emitter saturation intensities in each sample is presented in Fig.\ref{fig:stats}.  In sample SP1, without the SIL, 396 emitters were measured and the resulting distribution is shown in Fig.\ref{fig:stats}(a). The mean of the distribution is given as $\mu_{\mathrm{SP1}}$=\SI{68}{kcps}. Although a similar mean intensity to the NV centre in diamond, the variance in the saturation intensities differs significantly to what is seen in the NV centre in diamond where a narrow distribution is often reported \cite{Babinec2010a}, which can be correlated with the NV orientation. We attribute the larger variation in emitter brightness to differences in the internal energy levels between emitters which periodically suppress emission on different timescales. For example, a metastable shelving state \cite{Bishop2020} evident in the photon statistics of nearly all emitters, leads to pump power dependent bunching of emission on timescales of tens to hundreds of nanoseconds. Additionally, some emitters show telegraph noise which is observed on timescales of seconds, but may also be present on millisecond or microsecond scales. The relationship between these instabilities and the time-averaged intensity will be the subject of a future study. However, it is notable that the predicted photon CE should lead to a saturated emission intensity at least an order of magnitude greater than we measure, based on the extrapolated zero-pump power lifetime $t_1=\SI{4.6}{\nano\second}$ as determined from fitting the second order correlation measurements in Fig.\ref{fig:stats}(f).

In SP1, 396 emitters were sampled over an area of $\SI{150}{\micro\meter}$ by $\SI{150}{\micro\meter}$, corresponding to an area of $\SI{57}{\micro\meter^2}$ per emitter. Due to the magnification of the SIL, where the displacement in X and Y of our imaging system leads to a displacement of $X/n_{\mathrm{SIL}}$ and $Y/n_{\mathrm{SIL}}$ on the sample surface respectively, we ensure we sample a similar number of emitters per unit area scanned for SP2 and SP3. 63 and 70 emitters are measured for SP2 and SP3 respectively, with the corresponding distribution in $I_{\infty}$ shown in in Fig.\ref{fig:stats}(b,c).

The mean of the saturated intensity $I_{\infty}$ distribution in each sample is shown as the dashed vertical lines in Fig.\ref{fig:stats}(a-c). For SP2 and SP3, the mean $I_{\infty}$ is $\SI{211}{}$ and $\SI{285}{kcps}$, respectively. This corresponds to an enhancement factor of $\SI{3.1}{}$ and $\SI{4.2}{}$, relative to SP1. Assuming that the mean of the distribution for SP1 represents a CE of $\SI{4.6}{\percent}$, as determined from the analytical analysis, the projected CE for SP2 and SP3 can be estimated to be \SI{14.3}{\percent} and \SI{19.3}{\percent}. We note that the highest detected count rate is measured is an emitter in SP3 with $I_{\infty}=\SI{743\pm4}{kcps}$. We show data for this emitter in Fig.\ref{fig:stats}, labelled in (c) with a star. Pump power dependent auto-correlation data in (f) shows the presence of anti-bunching near time zero, in combination with longer timescale bunching at high pump powers. A fit to the $\SI{4}{\micro\meter}$ measurement, not shown in the figure, illustrates the single photon purity, with $g^{(2)}(0)=0.12$. In the experimental data we have removed an artefact at 8-\SI{30}{\nano\second} caused by optical cross-talk in our interferometer.

\subsection{Evanescent coupling}
To investigate the observed CE dependence on the polymer thickness we return to FDTD. Using the aforementioned simulation environment the electric field from a single dipole embedded in the middle of the AlN is captured on the surface of a box of monitors enclosing the emitter and spanning the dielectric stack. The near-field electric field on the XY monitor in the ZrO\textsubscript{2} layer is projected into the far field, where the flux is integrated over the solid angles of the collection optics to determine the collected power. The CE is defined as the collected power normalised by the power injected into the simulation via the single dipole. 

\begin{figure}
\centering
\includegraphics{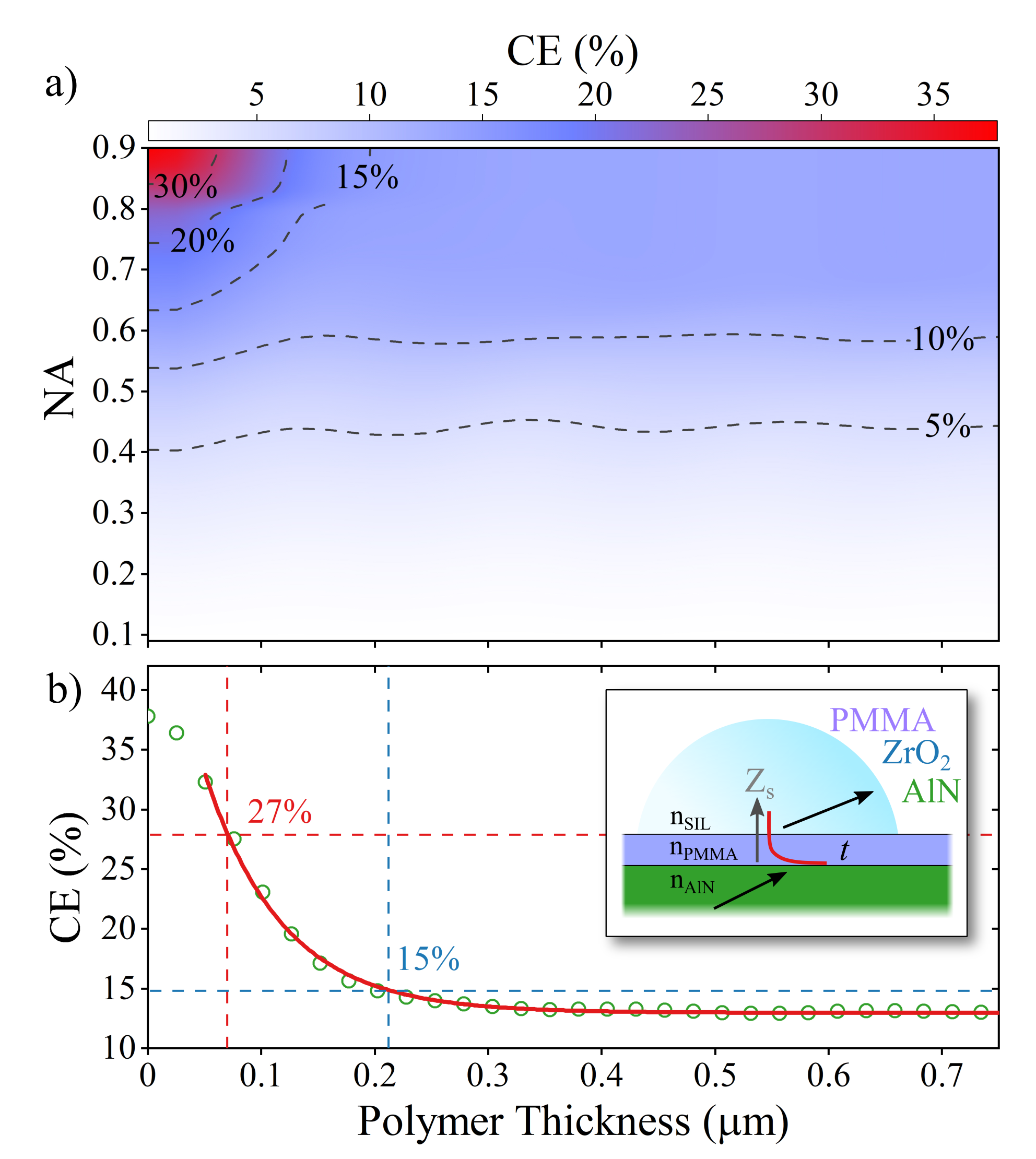}
\caption{Simulated CE demonstrating evanescent coupling through a thin polymer layer. a) CE as a function of the collection free space NA and the thickness $d$ of the PMMA polymer. b) CE as a function of the polymer thickness for NA$=\SI{0.9}{}$.}
\label{fig:fdtd}
\end{figure}

The CE as a function of the polymer thickness $d$ and NA of the imaging system is presented in Fig.4. The simulation shows that for $d>\SI{300}{\nano\meter}$ the CE is constant and is well below the expected CE for an ideal system with NA$=\SI{0.9}{}$, as determined in Fig.\ref{fig:intro}. This is due to total internal reflection at the AlN-PMMA interface at angles greater than $\SI{44}{\degree}$, limiting the effective collection NA to 0.69.

The greatest gain in CE can be made by combining a high NA lens with a thin PMMA layer. The CE for a NA of 0.9 as a function of the polymer thickness is presented in Fig.\ref{fig:fdtd}(b). We attribute the increased CE for polymer thickness less than $\SI{200}{\nano\meter}$ to evanescent coupling through the polymer for light incident at angles greater than $\SI{44}{}$ degrees. The evanescent wave in the polymer is converted back to a propagating wave at the polymer-SIL interface and collected by our imaging system, a phenomenon known as frustrated total internal reflection. The electric field in the polymer decays with the distance $z$ away from the interface (in the absence of the SIL) in the form,

\begin{equation}
    E(z)=E(s)e^{-2z/d} \quad\mathrm{and,}\quad  p = \frac{\lambda_{\mathrm{Air}}/n_{\mathrm{AlN}}}{2\pi\sqrt{\left(\frac{n_{\mathrm{AlN}}}{n_{\mathrm{PMMA}}}\right)^2\sin^2(\theta)-1}}
    \label{eq:evan}
\end{equation}

where $E(z)$ is the electric field normal to the interface, $E(s)$ is the electric field at the surface, $\theta$ is the incidence angle of the light, $\lambda_0$ is the wavelength of the light in a vacuum and $p$ is the penetration depth of the evanescent wave. 

Due to the angle dependence of the coupling through the polymer, we consider the two limiting angles of the system, just beyond the angle of TIR, $\theta_{\mathrm{TIR}}=\SI{45}{\degree}$, and the NA of the imaging system, $\theta_{\mathrm{NA}}=\SI{65}{\degree}$. The penetration depths as determined using Eq.\ref{eq:evan} are given as $p_{\mathrm{TIR}}=\SI{228}{\nano\meter}$ and $p_{\mathrm{NA}}=\SI{53}{\nano\meter}$, respectively. The determined penetration depths agree well with the simulated thickness dependent collection efficiency in Fig.\ref{fig:fdtd}, where for $d<\SI{200}{\nano\meter}$ the CE is increased above the vanishing value at infinite polymer thickness. A fit to the data in Fig.\ref{fig:fdtd}(b) using an offset exponential is used to interpolate the expected CE for SP2 and SP3. The determined CE values of $\SI{15}{}$ and $\SI{27}{\percent}$ agree well with the projected values from the statistical measurements.

\section{Conclusion}
The photon collection efficiency of quantum emitters in a AlN-on-sapphire substrate was enhanced by engineering the thickness of a PMMA adhesion layer under a hemispherical solid immersion lens. An automated emitter finding routine was used to determine the collection efficiency enhancement of an ensemble. We demonstrate experimentally and numerically that at high NA and low PMMA thickness evanescent coupling through the polymer can maximise CE. A collection efficiency of $\SI{19.3}{\percent}$ is inferred for a polymer thickness of \SI{70}{\nano\meter}, in good agreement with numerical simulations. The significant increase in collection efficiency from these room temperature quantum emitters is a step towards efficient off-the-shelf room temperature quantum light sources.

\section*{Acknowledgements}

We acknowledge financial support provided by EPSRC via Grant No. EP/T017813/1 and the European Union's H2020 Marie Curie ITN project LasIonDef (GA No. 956387). Device processing was carried out in the cleanroom of the ERDF-funded Institute for Compound Semiconductors (ICS) at Cardiff University. 






\bibliography{library}

\providecommand{\latin}[1]{#1}
\makeatletter
\providecommand{\doi}
  {\begingroup\let\do\@makeother\dospecials
  \catcode`\{=1 \catcode`\}=2 \doi@aux}
\providecommand{\doi@aux}[1]{\endgroup\texttt{#1}}
\makeatother
\providecommand*\mcitethebibliography{\thebibliography}
\csname @ifundefined\endcsname{endmcitethebibliography}
  {\let\endmcitethebibliography\endthebibliography}{}
\begin{mcitethebibliography}{23}
\providecommand*\natexlab[1]{#1}
\providecommand*\mciteSetBstSublistMode[1]{}
\providecommand*\mciteSetBstMaxWidthForm[2]{}
\providecommand*\mciteBstWouldAddEndPuncttrue
  {\def\EndOfBibitem{\unskip.}}
\providecommand*\mciteBstWouldAddEndPunctfalse
  {\let\EndOfBibitem\relax}
\providecommand*\mciteSetBstMidEndSepPunct[3]{}
\providecommand*\mciteSetBstSublistLabelBeginEnd[3]{}
\providecommand*\EndOfBibitem{}
\mciteSetBstSublistMode{f}
\mciteSetBstMaxWidthForm{subitem}{(\alph{mcitesubitemcount})}
\mciteSetBstSublistLabelBeginEnd
  {\mcitemaxwidthsubitemform\space}
  {\relax}
  {\relax}

\bibitem[Santori \latin{et~al.}(2002)Santori, Fattal, Vu{\v{c}}kovi{\'{c}},
  Solomon, and Yamamoto]{Santori2002}
Santori,~C.; Fattal,~D.; Vu{\v{c}}kovi{\'{c}},~J.; Solomon,~G.~S.; Yamamoto,~Y.
  {Indistinguishable photons from a single-photon device}. \emph{Nature}
  \textbf{2002}, \emph{419}, 594--597\relax
\mciteBstWouldAddEndPuncttrue
\mciteSetBstMidEndSepPunct{\mcitedefaultmidpunct}
{\mcitedefaultendpunct}{\mcitedefaultseppunct}\relax
\EndOfBibitem
\bibitem[Somaschi \latin{et~al.}(2016)Somaschi, Giesz, {De Santis}, Loredo,
  Almeida, Hornecker, Portalupi, Grange, Ant{\'{o}}n, Demory, G{\'{o}}mez,
  Sagnes, Lanzillotti-Kimura, Lema{\'{i}}tre, Auffeves, White, Lanco, and
  Senellart]{Somaschi2016}
Somaschi,~N. \latin{et~al.}  {Near-optimal single-photon sources in the solid
  state}. \emph{Nat. Photonics} \textbf{2016}, \emph{10}, 340--345\relax
\mciteBstWouldAddEndPuncttrue
\mciteSetBstMidEndSepPunct{\mcitedefaultmidpunct}
{\mcitedefaultendpunct}{\mcitedefaultseppunct}\relax
\EndOfBibitem
\bibitem[Babinec \latin{et~al.}(2010)Babinec, Hausmann, Khan, Zhang, Maze,
  Hemmer, and Lon{\v{c}}ar]{Babinec2010}
Babinec,~T.~M.; Hausmann,~B. J.~M.; Khan,~M.; Zhang,~Y.; Maze,~J.~R.;
  Hemmer,~P.~R.; Lon{\v{c}}ar,~M. {A diamond nanowire single-photon source}.
  \emph{Nat. Nanotechnol.} \textbf{2010}, \emph{5}, 195--199\relax
\mciteBstWouldAddEndPuncttrue
\mciteSetBstMidEndSepPunct{\mcitedefaultmidpunct}
{\mcitedefaultendpunct}{\mcitedefaultseppunct}\relax
\EndOfBibitem
\bibitem[Claudon \latin{et~al.}(2010)Claudon, Bleuse, Malik, Bazin, Jaffrennou,
  Gregersen, Sauvan, Lalanne, and G{\'{e}}rard]{Claudon2010}
Claudon,~J.; Bleuse,~J.; Malik,~N.~S.; Bazin,~M.; Jaffrennou,~P.;
  Gregersen,~N.; Sauvan,~C.; Lalanne,~P.; G{\'{e}}rard,~J.-M. {A highly
  efficient single-photon source based on a quantum dot in a photonic
  nanowire}. \emph{Nat. Photonics} \textbf{2010}, \emph{4}, 174--177\relax
\mciteBstWouldAddEndPuncttrue
\mciteSetBstMidEndSepPunct{\mcitedefaultmidpunct}
{\mcitedefaultendpunct}{\mcitedefaultseppunct}\relax
\EndOfBibitem
\bibitem[Noda \latin{et~al.}(2007)Noda, Fujita, and Asano]{Noda2007}
Noda,~S.; Fujita,~M.; Asano,~T. {Spontaneous-emission control by photonic
  crystals and nanocavities}. \emph{Nat. Photonics} \textbf{2007}, \emph{1},
  449--458\relax
\mciteBstWouldAddEndPuncttrue
\mciteSetBstMidEndSepPunct{\mcitedefaultmidpunct}
{\mcitedefaultendpunct}{\mcitedefaultseppunct}\relax
\EndOfBibitem
\bibitem[Lodahl \latin{et~al.}(2015)Lodahl, Mahmoodian, and
  Stobbe]{Lodahl2015a}
Lodahl,~P.; Mahmoodian,~S.; Stobbe,~S. {Interfacing single photons and single
  quantum dots with photonic nanostructures}. \emph{Rev. Mod. Phys.}
  \textbf{2015}, \emph{87}, 347--400\relax
\mciteBstWouldAddEndPuncttrue
\mciteSetBstMidEndSepPunct{\mcitedefaultmidpunct}
{\mcitedefaultendpunct}{\mcitedefaultseppunct}\relax
\EndOfBibitem
\bibitem[Kuznetsov \latin{et~al.}(2016)Kuznetsov, Miroshnichenko, Brongersma,
  Kivshar, and Luk'yanchuk]{Kuznetsov2016}
Kuznetsov,~A.~I.; Miroshnichenko,~A.~E.; Brongersma,~M.~L.; Kivshar,~Y.~S.;
  Luk'yanchuk,~B. {Optically resonant dielectric nanostructures}. \emph{Science
  (80-. ).} \textbf{2016}, \emph{354}\relax
\mciteBstWouldAddEndPuncttrue
\mciteSetBstMidEndSepPunct{\mcitedefaultmidpunct}
{\mcitedefaultendpunct}{\mcitedefaultseppunct}\relax
\EndOfBibitem
\bibitem[Mignuzzi \latin{et~al.}(2019)Mignuzzi, Vezzoli, Horsley, Barnes,
  Maier, and Sapienza]{Mignuzzi2019}
Mignuzzi,~S.; Vezzoli,~S.; Horsley,~S.~A.; Barnes,~W.~L.; Maier,~S.~A.;
  Sapienza,~R. {Nanoscale Design of the Local Density of Optical States}.
  \emph{Nano Lett.} \textbf{2019}, \emph{19}, 1613--1617\relax
\mciteBstWouldAddEndPuncttrue
\mciteSetBstMidEndSepPunct{\mcitedefaultmidpunct}
{\mcitedefaultendpunct}{\mcitedefaultseppunct}\relax
\EndOfBibitem
\bibitem[Tomm \latin{et~al.}(2021)Tomm, Javadi, Antoniadis, Najer, L{\"{o}}bl,
  Korsch, Schott, Valentin, Wieck, Ludwig, and Warburton]{Tomm2021}
Tomm,~N.; Javadi,~A.; Antoniadis,~N.~O.; Najer,~D.; L{\"{o}}bl,~M.~C.;
  Korsch,~A.~R.; Schott,~R.; Valentin,~S.~R.; Wieck,~A.~D.; Ludwig,~A.;
  Warburton,~R.~J. {A bright and fast source of coherent single photons}.
  \emph{Nat. Nanotechnol.} \textbf{2021}, \emph{16}, 399--403\relax
\mciteBstWouldAddEndPuncttrue
\mciteSetBstMidEndSepPunct{\mcitedefaultmidpunct}
{\mcitedefaultendpunct}{\mcitedefaultseppunct}\relax
\EndOfBibitem
\bibitem[Dolan \latin{et~al.}(2010)Dolan, Hughes, Grazioso, Patton, and
  Smith]{Dolan2010}
Dolan,~P.~R.; Hughes,~G.~M.; Grazioso,~F.; Patton,~B.~R.; Smith,~J.~M.
  {Femtoliter tunable optical cavity arrays}. \emph{Opt. Lett.} \textbf{2010},
  \emph{35}, 3556\relax
\mciteBstWouldAddEndPuncttrue
\mciteSetBstMidEndSepPunct{\mcitedefaultmidpunct}
{\mcitedefaultendpunct}{\mcitedefaultseppunct}\relax
\EndOfBibitem
\bibitem[Born and Wolf(1999)Born, and Wolf]{Born1999}
Born,~M.; Wolf,~E. \emph{Princ. Opt. Electromagn. Theory Propag. Interf. Diffr.
  Light}; 1999\relax
\mciteBstWouldAddEndPuncttrue
\mciteSetBstMidEndSepPunct{\mcitedefaultmidpunct}
{\mcitedefaultendpunct}{\mcitedefaultseppunct}\relax
\EndOfBibitem
\bibitem[Vamivakas \latin{et~al.}(2007)Vamivakas, Atat{\"{u}}re, Dreiser,
  Yilmaz, Badolato, Swan, Goldberg, Imamoǧlu, and
  {\"{U}}nl{\"{u}}]{A.N.Vamivakas2007}
Vamivakas,~A.~N.; Atat{\"{u}}re,~M.; Dreiser,~J.; Yilmaz,~S.~T.; Badolato,~A.;
  Swan,~A.~K.; Goldberg,~B.~B.; Imamoǧlu,~A.; {\"{U}}nl{\"{u}},~M.~S. {Strong
  extinction of a far-field laser beam by a single quantum dot}. \emph{Nano
  Lett.} \textbf{2007}, \emph{7}, 2892--2896\relax
\mciteBstWouldAddEndPuncttrue
\mciteSetBstMidEndSepPunct{\mcitedefaultmidpunct}
{\mcitedefaultendpunct}{\mcitedefaultseppunct}\relax
\EndOfBibitem
\bibitem[Chen \latin{et~al.}(2018)Chen, Zopf, Keil, Ding, and
  Schmidt]{Chen2018}
Chen,~Y.; Zopf,~M.; Keil,~R.; Ding,~F.; Schmidt,~O.~G. {Highly-efficient
  extraction of entangled photons from quantum dots using a broadband optical
  antenna}. \emph{Nat. Commun.} \textbf{2018}, \emph{9}, 1--7\relax
\mciteBstWouldAddEndPuncttrue
\mciteSetBstMidEndSepPunct{\mcitedefaultmidpunct}
{\mcitedefaultendpunct}{\mcitedefaultseppunct}\relax
\EndOfBibitem
\bibitem[Barnes \latin{et~al.}(2002)Barnes, Bj{\"{o}}rk, G{\'{e}}rard, Jonsson,
  Wasey, Worthing, and Zwiller]{Barnes2002}
Barnes,~W.; Bj{\"{o}}rk,~G.; G{\'{e}}rard,~J.; Jonsson,~P.; Wasey,~J.;
  Worthing,~P.; Zwiller,~V. {Solid-State Single Photon Sources : Light
  Collection Strategies}. \emph{Eur. Phys. J. D - At. Mol. Opt. Plasma Phys.}
  \textbf{2002}, \emph{210}, 197--210\relax
\mciteBstWouldAddEndPuncttrue
\mciteSetBstMidEndSepPunct{\mcitedefaultmidpunct}
{\mcitedefaultendpunct}{\mcitedefaultseppunct}\relax
\EndOfBibitem
\bibitem[Serrels \latin{et~al.}(2008)Serrels, Ramsay, Dalgarno, Gerardot,
  O'Connor, Hadfield, Warburton, and Reid]{Serrels2008}
Serrels,~K.~A.; Ramsay,~E.; Dalgarno,~P.~A.; Gerardot,~B.; O'Connor,~J.~A.;
  Hadfield,~R.~H.; Warburton,~R.~J.; Reid,~D.~T. {Solid immersion lens
  applications for nanophotonic devices}. \emph{J. Nanophotonics}
  \textbf{2008}, \emph{2}, 021854\relax
\mciteBstWouldAddEndPuncttrue
\mciteSetBstMidEndSepPunct{\mcitedefaultmidpunct}
{\mcitedefaultendpunct}{\mcitedefaultseppunct}\relax
\EndOfBibitem
\bibitem[Zhu and Oliver(2016)Zhu, and Oliver]{Zhu2016a}
Zhu,~T.; Oliver,~R.~A. {Nitride quantum light sources}. \emph{EPL (Europhysics
  Lett.} \textbf{2016}, \emph{113}, 38001\relax
\mciteBstWouldAddEndPuncttrue
\mciteSetBstMidEndSepPunct{\mcitedefaultmidpunct}
{\mcitedefaultendpunct}{\mcitedefaultseppunct}\relax
\EndOfBibitem
\bibitem[Berhane \latin{et~al.}(2017)Berhane, Jeong, Bodrog, Fiedler,
  Schr{\"{o}}der, Trivi{\~{n}}o, Palacios, Gali, Toth, Englund, and
  Aharonovich]{Berhane2017}
Berhane,~A.~M.; Jeong,~K.-Y.; Bodrog,~Z.; Fiedler,~S.; Schr{\"{o}}der,~T.;
  Trivi{\~{n}}o,~N.~V.; Palacios,~T.; Gali,~A.; Toth,~M.; Englund,~D.;
  Aharonovich,~I. {Bright Room-Temperature Single-Photon Emission from Defects
  in Gallium Nitride}. \emph{Adv. Mater.} \textbf{2017}, \emph{29},
  1605092\relax
\mciteBstWouldAddEndPuncttrue
\mciteSetBstMidEndSepPunct{\mcitedefaultmidpunct}
{\mcitedefaultendpunct}{\mcitedefaultseppunct}\relax
\EndOfBibitem
\bibitem[Zhou \latin{et~al.}(2018)Zhou, Wang, Rasmita, Kim, Berhane, Bodrog,
  Adamo, Gali, Aharonovich, and Gao]{Zhou2018}
Zhou,~Y.; Wang,~Z.; Rasmita,~A.; Kim,~S.; Berhane,~A.; Bodrog,~Z.; Adamo,~G.;
  Gali,~A.; Aharonovich,~I.; Gao,~W.-b. {Room temperature solid-state quantum
  emitters in the telecom range}. \emph{Sci. Adv.} \textbf{2018}, \emph{4},
  eaar3580\relax
\mciteBstWouldAddEndPuncttrue
\mciteSetBstMidEndSepPunct{\mcitedefaultmidpunct}
{\mcitedefaultendpunct}{\mcitedefaultseppunct}\relax
\EndOfBibitem
\bibitem[Bishop \latin{et~al.}(2020)Bishop, Hadden, Alzahrani, Hekmati,
  Huffaker, Langbein, and Bennett]{Bishop2020}
Bishop,~S.~G.; Hadden,~J.~P.; Alzahrani,~F.~D.; Hekmati,~R.; Huffaker,~D.~L.;
  Langbein,~W.~W.; Bennett,~A.~J. {Room-Temperature Quantum Emitter in Aluminum
  Nitride}. \emph{ACS Photonics} \textbf{2020}, acsphotonics.0c00528\relax
\mciteBstWouldAddEndPuncttrue
\mciteSetBstMidEndSepPunct{\mcitedefaultmidpunct}
{\mcitedefaultendpunct}{\mcitedefaultseppunct}\relax
\EndOfBibitem
\bibitem[Wan \latin{et~al.}(2020)Wan, Lu, Chen, Walsh, Trusheim, {De Santis},
  Bersin, Harris, Mouradian, Christen, Bielejec, and Englund]{Wan2020}
Wan,~N.~H.; Lu,~T.~J.; Chen,~K.~C.; Walsh,~M.~P.; Trusheim,~M.~E.; {De
  Santis},~L.; Bersin,~E.~A.; Harris,~I.~B.; Mouradian,~S.~L.; Christen,~I.~R.;
  Bielejec,~E.~S.; Englund,~D. {Large-scale integration of artificial atoms in
  hybrid photonic circuits}. \emph{Nature} \textbf{2020}, \emph{583},
  226--231\relax
\mciteBstWouldAddEndPuncttrue
\mciteSetBstMidEndSepPunct{\mcitedefaultmidpunct}
{\mcitedefaultendpunct}{\mcitedefaultseppunct}\relax
\EndOfBibitem
\bibitem[Bishop \latin{et~al.}(2022)Bishop, Hadden, Hekmati, Cannon, Langbein,
  and Bennett]{Bishop2022}
Bishop,~S.~G.; Hadden,~J.~P.; Hekmati,~R.; Cannon,~J.~K.; Langbein,~W.~W.;
  Bennett,~A.~J. {Enhanced light collection from a gallium nitride color center
  using a near index-matched solid immersion lens}. \emph{Appl. Phys. Lett.}
  \textbf{2022}, \emph{120}, 114001\relax
\mciteBstWouldAddEndPuncttrue
\mciteSetBstMidEndSepPunct{\mcitedefaultmidpunct}
{\mcitedefaultendpunct}{\mcitedefaultseppunct}\relax
\EndOfBibitem
\bibitem[Babinec \latin{et~al.}(2010)Babinec, Hausmann, Khan, Zhang, Maze,
  Hemmer, and Lon{\v{c}}ar]{Babinec2010a}
Babinec,~T.~M.; Hausmann,~B.~J.; Khan,~M.; Zhang,~Y.; Maze,~J.~R.;
  Hemmer,~P.~R.; Lon{\v{c}}ar,~M. {A diamond nanowire single-photon source}.
  \emph{Nat. Nanotechnol.} \textbf{2010}, \emph{5}, 195--199\relax
\mciteBstWouldAddEndPuncttrue
\mciteSetBstMidEndSepPunct{\mcitedefaultmidpunct}
{\mcitedefaultendpunct}{\mcitedefaultseppunct}\relax
\EndOfBibitem
\end{mcitethebibliography}

\end{document}